\documentclass[reprint,showpacs,amsmath,amssymb,showkeys,aps,prl]{revtex4-1}
\usepackage{natbib}
\usepackage{graphicx}
\usepackage{dcolumn}
\usepackage{bm}
\usepackage{mathtools}
\pdfoutput=1
\usepackage{color}
\usepackage[pdftex]{hyperref}
\hypersetup{colorlinks=true,citecolor=blue,linkcolor=blue,urlcolor=blue}
\usepackage[all]{hypcap}

\newcommand\underrel[2]{\mathrel{\mathop{#2}\limits_{#1}}}
\definecolor{green}{rgb}{0,0.5,0}

\begin{document}

\title{Metadynamics of paths}

\author{Davide Mandelli$^{a}$, Barak Hirshberg$^{b,c}$, Michele Parrinello$^{a,b,c,*}$}

\affiliation{
$^a$Atomistic Simulations, Italian Institute of Technology, via Morego 30, 16163 Genova, Italy \\
$^b$Department of Chemistry and Applied Biosciences, ETH Zurich, 8092 Zurich, Switzerland \\
$^c$Institute of Computational Sciences, Universit\`a della Svizzera Italiana, 6900 Lugano, Switzerland
}

\email[]{michele.parrinello@phys.chem.ethz.ch}
\date{\today}

\begin{abstract}
We present a method to sample reactive pathways via biased molecular dynamics simulations in trajectory space. We show that the use of enhanced sampling techniques enables unconstrained exploration of multiple reaction routes. Time correlation functions are conveniently computed via reweighted averages along a single trajectory and kinetic rates are accessed at no additional cost. These abilities are illustrated analyzing a model potential and the umbrella inversion of NH$_3$ in water. The algorithm allows a parallel implementation and promises to be a powerful tool for the study of rare events.
\end{abstract}

\pacs{}
\maketitle

%%%%%% MAIN TEXT %%%%%%
Molecular dynamics (MD) simulations have become an invaluable tool in many branches of science. While experiments generally access only spatially and time averaged quantities, atomically detailed MD simulations allow tracking in real time the microscopic mechanisms underlying complex phenomena. Nevertheless, there is a large class of problems where a straightforward application of MD simulations is impractical. Important examples are crystal nucleation, slow diffusion in solids, chemical reactions and conformational changes of large molecules. In all these cases, the presence of large free energy barriers leads to impractically long computational times. Therefore, it is necessary to design efficient algorithms able to accelerate phase space exploration.

A vast number of such methods have been proposed. Here we focus on Metadynamics~\cite{Laio2002} (MetaD) that has recently gained great popularity. In MetaD, as in other similar methods,  sampling is accelerated by the addition to the Hamiltonian of an external potential, also referred to as bias. However, the addition of this potential changes the natural dynamics of the system and only using an especially engineered bias some dynamical properties can be retrieved~\cite{Tiwary2013,Wu2014,Rosta2015,Wu2016}. In a more ambitious effort, Donati {\it et al.}~\cite{Donati2017,Donati2018} have described a general method to recover dynamical properties from biased trajectories. However, the procedure suggested is prone to numerical instabilities.

Other researchers have taken a different point of view and direct attention has been focused on reactive paths (RPs) and their sampling~\cite{Elber1987,Olender1996,Passerone2001,Lee2017,Pratt1986,Dellago1998a,Fujisaki2010}. A successful and widely used path-based method is transition path sampling (TPS) that is a Monte Carlo procedure for harvesting RPs that connect two a priori known metastable states~\cite{Dellago1998a}. The theoretical underpinning of this and similar approaches is the Onsager-Machlup (OM) action that determines the path probability distribution, as we shall discuss below. While highly successful, applications of TPS are met with some difficulties. The initial and final states need to be known beforehand, along with at least one RP connecting them. The computation of rate constants can also be time consuming~\cite{Dellago1998a,VanErp2003,Moroni2004}. Furthermore, if different pathways are possible (see figure~\ref{fig1}) one encounters sampling problems such as path trapping in the vicinity of the original guess~\cite{Vlugt2001,Borrero2016,Bolhuis2018a}.

In this letter we combine the power of MetaD and path-based methods and show that one can harvest reactive trajectories without choosing a final state and opening the possibility of exploring multiple pathways in a single run. Although we apply an external bias, equilibrium time correlation functions can be straightforwardly obtained with reweighting procedures that do not encounter numerical problems. In the following, we briefly review the theory and formalism behind the algorithm and then present two applications. First to a model system, meant to show sampling of multiple reactive paths in one simulation. The second demonstrates the utility of the method in obtaining time correlation functions and kinetic rates in the realistic case of ammonia in water.

The problem of interest here is the time evolution of a system coupled to a thermal bath at temperature $T$. Onsager and Machlup~\cite{Onsager1953} have shown that in the overdamped regime the probability of observing a trajectory $R(t)$ of duration $\tau$ is given by
\begin{equation}
\label{eq1}
P[R(t)]\propto e^{-S[R(t)]},
\end{equation}
where the OM action is defined as
\begin{equation}
\label{eq2}
S[R(t)]=\int_0^{\tau}\frac {1} {2\sigma^2}\left( \dot R(t)-\frac{F(t)}{m\nu} \right)^2{\rm d}t.
\end{equation}
Here, $m$ and $\dot R$ are the mass and velocity of the system and $F$ is the force acting on it while $\nu$ is a friction coefficient and $\sigma^2$=$2k_BT/m\nu$.

We will consider the dynamics of a molecular system composed of $M$ atoms, described by a $3M$-dimensional coordinate vector ${\bf R}$=$\{{\bf r}_j\}_{j=1,M}$. In numerical applications, a trajectory ${\bf R}(t)$ of duration $\tau$ is discretized into $N$ configurations ${\bf R}^n$ equally spaced in time and labelled by an index $n$=$1,2,\dots,N$, and the OM action~\eqref{eq2} becomes:
\begin{equation}
\label{eq3}
S=\sum_{n=1}^{N-1}\sum_{j=1}^M\frac{1}{2\sigma_j^2}\left(\frac{{\bf r}^{n+1}_j-{\bf r}_j^n}{\Delta t}-\frac{{\bf F}_j^n}{m_j\nu}\right)^2\Delta t.
\end{equation}
Here, $\Delta t$=$\tau/(N-1)$, $m_j$ is the mass of atom $j$, ${\bf F}_j^n$=$-\nabla_{{\bf r}_j^n}V({\bf R}^n)$ is the force acting on it in the $n$-th configuration while $\sigma_j^2$=$2k_BT/m_j\nu$ . Furthermore, we shall not consider one single trajectory but an ensemble of trajectories that start from an initial metastable state, thus, we shall draw the initial configuration ${\bf R}^1$ from the Boltzmann distribution $P({\bf R}^1)\propto e^{-\beta V({\bf R}^1)}$. Combining this with the OM probability~\eqref{eq1}, the probability of observing a discretized trajectory ${\bf R}^1\rightarrow{\bf R}^2\rightarrow\dots\rightarrow{\bf R}^N$ can then be expressed as
\begin{equation}
\label{eq4}
P[{\bf R}^1,{\bf R}^2,\dots,{\bf R}^N]\propto e^{-\beta V_{\rm eff}({\bf R}^1,{\bf R}^2,\dots,{\bf R}^N)},
\end{equation}
where $\beta$=$1/k_BT$ and
\begin{equation}
\label{eq5}
V_{\rm eff}=V({\bf R}^1)+\sum_{n=1}^{N-1}\sum_{j=1}^M\frac{K_j}{2}\left({\bf r}^{n+1}_j-{\bf r}_j^n-{\bf L}_j^n\right)^2,
\end{equation}
where we have defined the spring constant $K_j$=$\frac{m_j\nu}{2\Delta t}$ and the equilibrium length ${\bf L}_j^n$=$\frac{\Delta t}{m_j\nu}{\bf F}^n_j$.
\begin{figure}[tb]\label{fig0.polymer}
  \centering
  \includegraphics[width=0.8\columnwidth]{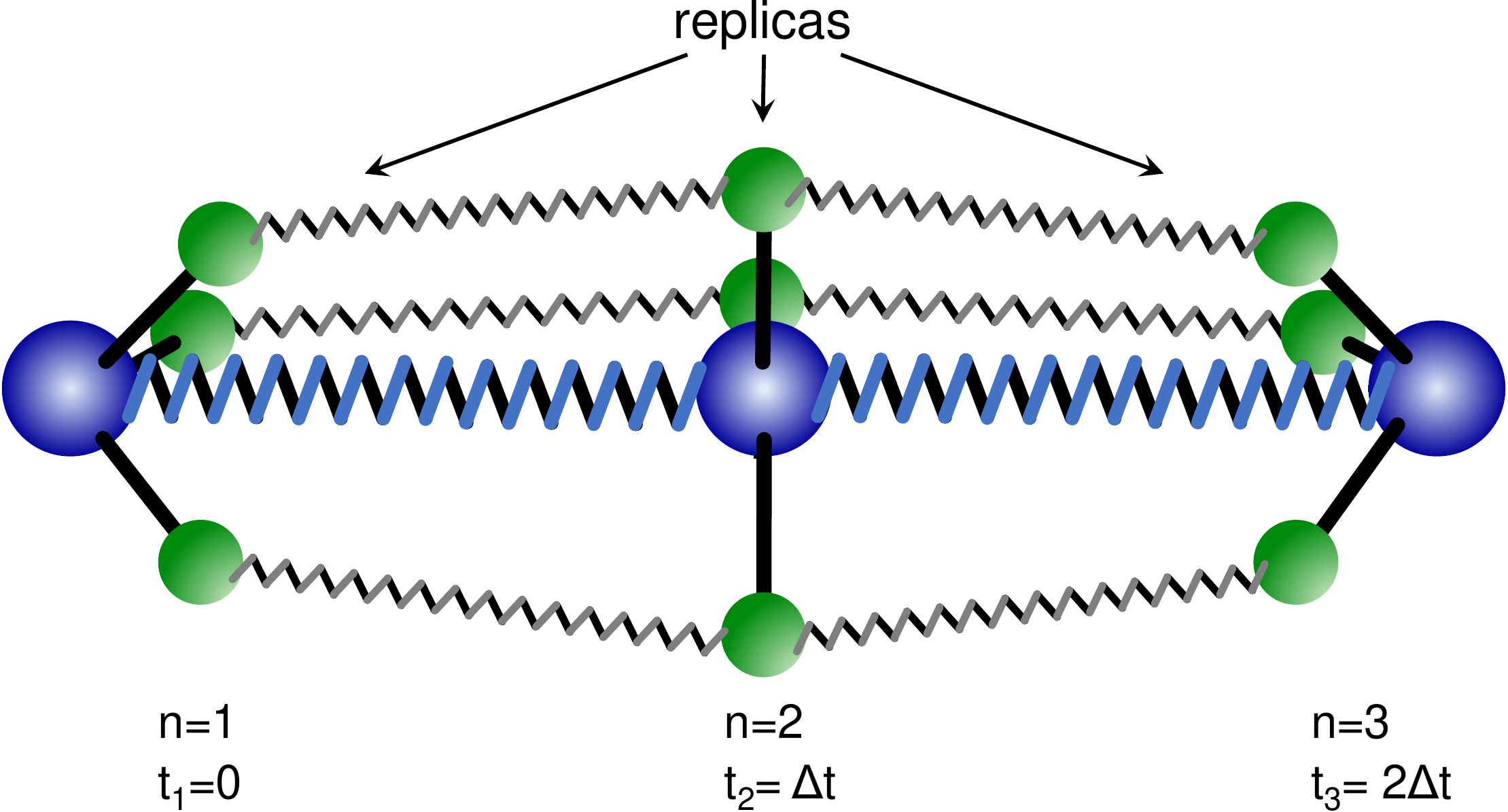}
  \caption{Polymer model representing a two-step path for the umbrella inversion of NH$_3$. This extreme discretization is only used for illustration purposes.}
\end{figure}

The effective potential $V_{\rm eff}$ can be interpreted as that of an open polymer of $N$ beads ${\bf R}^n$ corresponding to the configurations visited along the trajectory at times $t_{n}$=$(n-1)\Delta t$. The atoms in adjacent beads are linked by springs and the first bead feels the potential $V({\bf R}^1)$. This fictitious system is illustrated in figure~\ref{fig0.polymer} for the case of one ammonia molecule.

The observation to make here is that with these manipulations we have mapped a dynamical problem into a time independent polymer problem. Thus, one can sample $V_{\rm eff}({\bf R}^1,{\bf R}^2,\dots,{\bf R}^N)$ with standard methods. Here, we use Hamiltonian sampling as done for instance in path integral MD~\cite{Parrinello1984}. That is, we attribute to the polymer beads artificial masses, couple it to a thermostat and generate polymer trajectories. Assuming an ergodic behaviour, temporal averages over this fictitious dynamics are equivalent to ensemble averages. 

In order to evolve this dynamics we need to calculate the forces $-\nabla_{{\bf R}^n}V_{\rm eff}({\bf R}^1,{\bf R}^2,\dots,{\bf R}^N)$, which implies calculating the second derivatives of the physical potential $V({\bf R})$, since $V_{\rm eff}$ depends on the first derivatives via the terms ${\bf F}_j^n$ [see equation~\eqref{eq5}]. This would much worsen the scaling of the algorithm with system size. This consideration has discouraged other researchers from following a path similar to ours~\cite{Dellago1998a}. We get around this technical problem by using a finite difference formula that is illustrated in the Supplemental Material~\cite{SM}.\nocite{jorgensen1983,Chandler1987,Hockney,Martina1992,Putrino2000,Tribello2014,Kapil2016} Adopting this method, one time step in path space involves 3$\times$$N$ force evaluations. This has to be compared with the cost of $N$ MD steps needed to generate a new trajectory in standard simulations, which involve $N$ force evaluations. However, while the standard approach is intrinsically serial, the path approach has the advantage that it can be made highly parallel. Specifically, here we adopt the hyper-parallel scheme of Calhoun {\it et al.}~\cite{Calhoun1996} and implement the algorithm in the LAMMPS~\cite{Plimpton1995} suite of codes.

Sampling $V_{\rm eff}$ is not without problems. Complex systems are characterized by many different RPs and the ultimate goal of path-sampling algorithms is to sample all of them. Path trapping occurs when the algorithm is not able to locate other RPs than those close to the initial guess. This is an important problem addressed in our work. Our approach does not require an initial guess of the RP or knowledge of the final state. Instead, we use MetaD to sample different RPs, even when they are separated by high energy barriers. However, the MetaD bias changes the statistical weight of the trajectories sampled and this needs to be accounted for. This is done via the well tested and stable reweighting methods that have been developed in the MetaD literature~\cite{Bonomi2009,Branduardi2012,tiwary2015,Mones2016,Marinova2019,invernizzi2020,giberti2020}.

Once we have sampled a sufficient number of trajectories and their weights, we can calculate the dynamical properties of interest. Here we will focus on the correlation function introduced by Miller~\cite{Miller1974} to study the transitions from basin A to basin B
\begin{equation}
\label{eq6}
C(t)=\frac{\langle I_{\rm A}(0)I_{\rm B}(t)\rangle}{\langle I_{\rm A}(0)\rangle},
\end{equation}
where the characteristic function $I_{\rm X}(t)$ is 1 if at time $t$ the system is in basin X and 0 otherwise. $C(t)$ measures the probability for a system that is in A at time 0 to make a transition to B at time $t$. As Miller has shown~\cite{Miller1983}, in a rare event scenario
\begin{equation}
\label{eq7}
C(t)\underrel{t\to\infty}=k_{\rm AB}t,
\end{equation}
where $k_{\rm AB}$ is the phenomenological transition rate. Thus, in our method, the calculation of $C(t)$ is performed computing averages over the polymer configurations (see Supplemetal Material~\cite{SM}). The rate $k_{\rm AB}$ is then extracted from its asymptotic behaviour.
\begin{figure}[tb]\label{fig1}
  \centering
  \includegraphics[width=0.8\columnwidth]{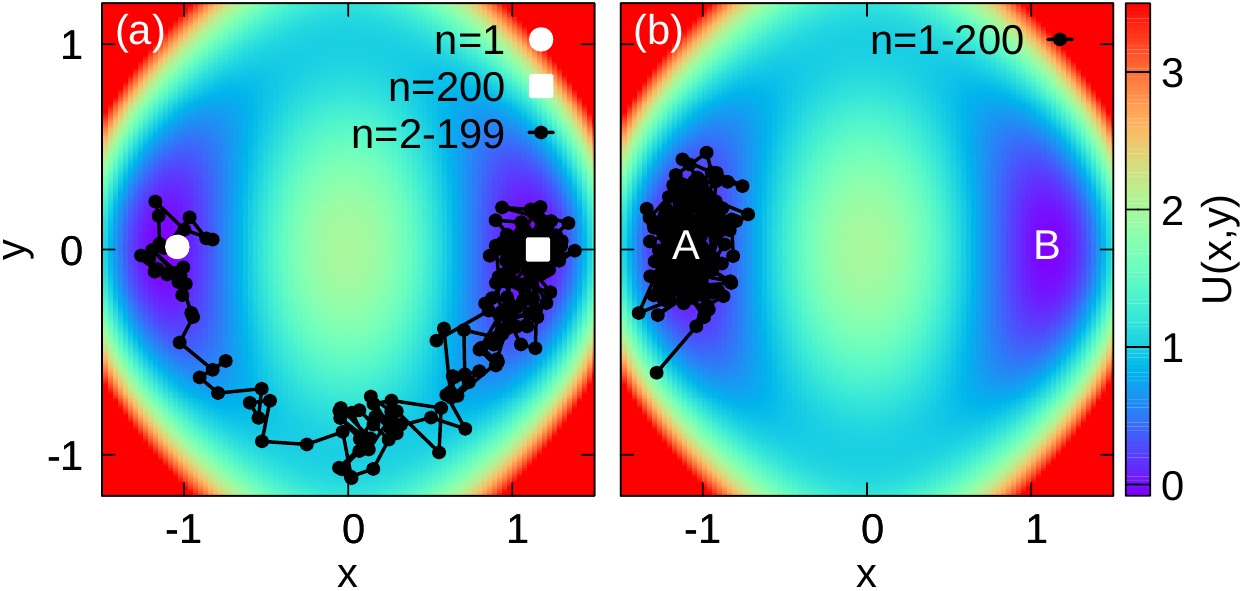}\\
  \vspace{0.1cm}
  \includegraphics[width=0.8\columnwidth]{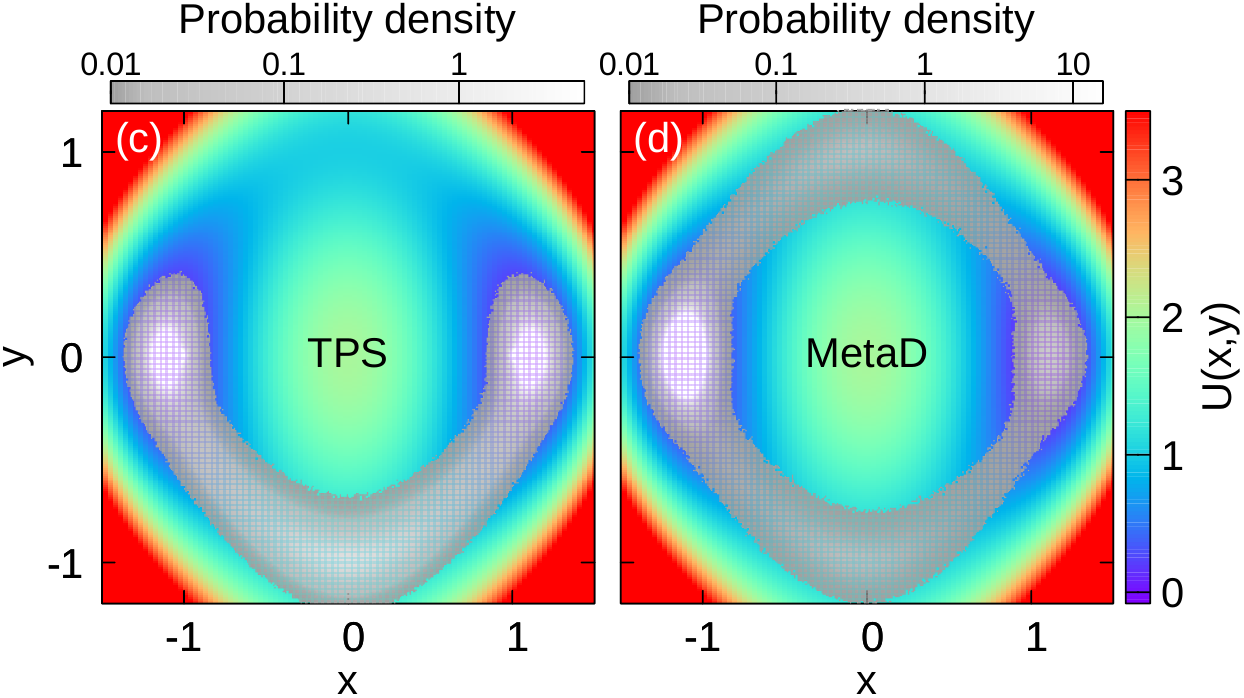}
  \caption{(a),(b) Polymer configurations representing (a) a reactive path crossing the lower saddle and (b) a non-reactive path crumpled in the left basin. A and B mark the two minima of the potential. (c),(d) Probability distribution of the positions of all beads obtained from (c) dynamical TPS and (d) MetaD in path space. The MetaD result is obtained without reweighting.}
\end{figure}

We will now illustrate the method with some applications. For the sake of streamlining the presentation we omit here most of the technical details. A full technical description of our runs can be found in the Supplemental Material~\cite{SM}. As a first test case, we consider the dynamics of a particle in the two dimensional double-well potential of figure~\ref{fig1}~\cite{Dellago1998a}. This model provides a simple example of a system with multiple RPs connecting metastable states. Simulations in path space are carried out using a polymer of size $N$=200 beads. In the first set of calculations, similar in spirit to TPS~\cite{Dellago1998a}, we fix with harmonic springs the initial and final positions in the left and right minimum respectively, thus restricting ourselves to the study of reactive paths. A representative path passing via the lower saddle is shown in figure~\ref{fig1}(a). If we sample the trajectories starting from this initial one the probability of sampling the upper saddle is vanishingly small due to the large potential energy that separates the two paths and only the lower paths will be explored [see figure~\ref{fig1}(c)]. This is one of the well known problems of TPS~\cite{Vlugt2001,Borrero2016,Bolhuis2018a}.

In order to overcome this difficulty, we first remove the constraint that the path should end in the right basin. If we do this and run an unbiased simulation as described above only crumpled trajectories localized in the initial basin are observed [see figure~\ref{fig1}(b)]. This reflects the physical fact that transitions between metastable states are rare events and therefore the probability of sampling RPs is very low. A possible way of observing RPs is to enhance trajectory sampling with the use of MetaD. MetaD is a rigorous procedure to enhance the fluctuations of selected degrees of freedom or collective variables (CVs)~\cite{Laio2002,Barducci2008,Dama2014}. In our case, since we want to sample trajectories that instead of remaining crumpled span the range from A to B, a natural choice is to use as CV the end-to-end distance of the polymer $d_{\rm e2e}=\lvert{\bf R}^N-{\bf R}^1\rvert$, thus enhancing the probability of sampling paths that go from A to B. The results of this MetaD biased simulations are shown in figure~\ref{fig1}(d), where it can be seen that both branches are equally sampled. The calculation of the transition rate for this system is reported in the Supplemental Material~\cite{SM}.
\begin{figure}[tb]\label{fig4.rateNH3}
  \centering
  \includegraphics[width=0.8\columnwidth]{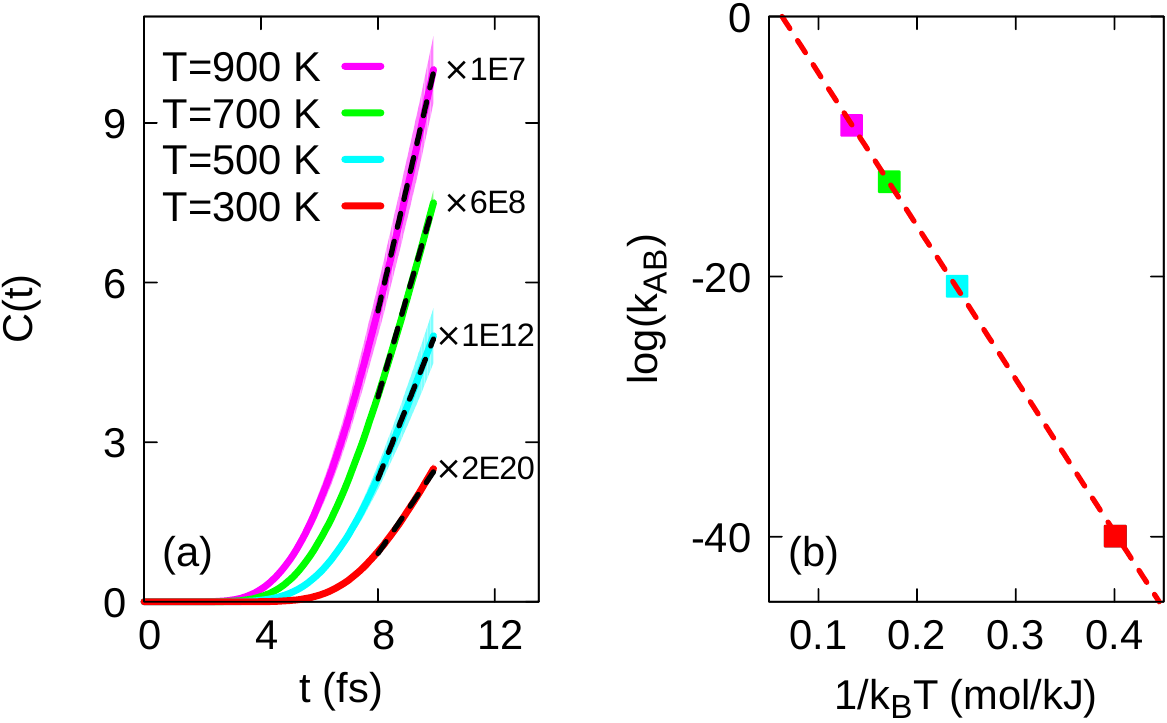}
  \caption{(a) The time correlation function computed at different temperatures. Dashed lines are fit to $k_{\rm AB}t+a$. Labels are scaling factors used for clarity of presentation. (b) Arrhenius plot of the rate constants. The red dashed line is a fit to $-\Delta E_{\rm fit}/k_BT+b$.}
\end{figure}

Having demonstrated the capabilities of our method in a simple model, we apply it to a more realistic case, namely, the umbrella inversion of NH$_3$ in water. In this transition, the nitrogen atom passes through the hydrogen plane to reach an equivalent and symmetric position. Thus, the process is conveniently described in terms of the oriented height $h$ of the NH$_3$ tetrahedron. Before tackling the NH$_3$ inversion in water we start by considering the same problem in vacuum. This study has a double purpose. On the one hand, it is another and more realistic problem on which to check our machinery, on the other, it allows us to understand the role of the solvent in the NH$_3$ inversion.

Here, we describe intra-molecular forces using an empirical model~\cite{Weismiller2010}. At equilibrium, $h$ takes values of $h^{\rm eq}\approx\pm0.4$ \AA, the two equivalent configurations being separated by a large barrier of $\Delta E\approx$120 kJ/mol$\approx$50 $k_BT$. Simulations in path space are performed using a polymer of size $N$=100. We check first that the size of the polymer is large enough to obtain converged results~\cite{SM}. We choose as before the generalized end-to-end distance $\Delta h_{\rm e2e}=(h^{N}-h^{1})$ as CV. Biased simulations are performed adopting OPES~\cite{invernizzi2020} that is an efficient and very recent evolution of MetaD. We do not impose any constraint on the polymer and compute $C(t)$ via equation~\eqref{eq6}. This is reported in figure~\ref{fig4.rateNH3}(a). In panel (b) we show the Arrhenius plot of the phenomenological rates $k_{\rm AB}$, showing the expected linear trend. A fit of the data yields an activation barrier of $\Delta E_{\rm fit}$=$118\pm1$ kJ/mol, in agreement with the exact value of $\Delta E\approx$120 kJ/mol for the force field adopted.
\begin{figure}[t]\label{fig5.dry-VS-wet}
  \centering
  \includegraphics[width=0.8\columnwidth]{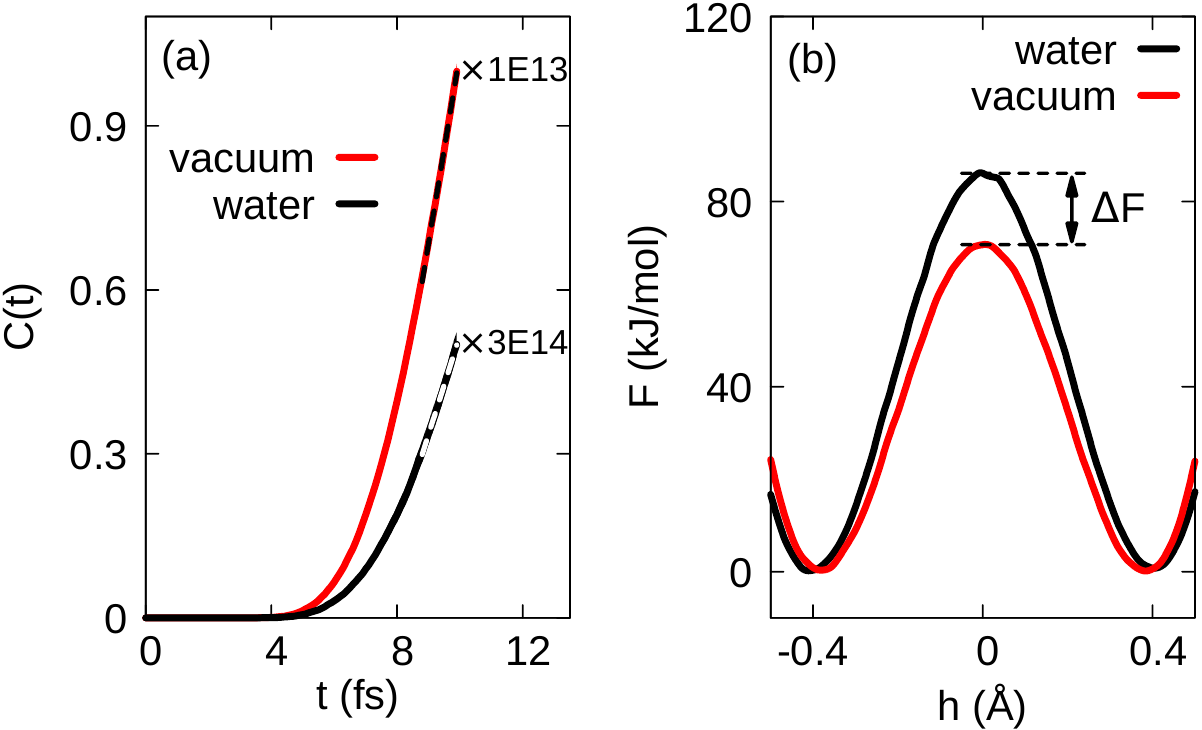}
  \caption{(a) The time correlation function computed with (black) and without water (red). Dashed lines are fit to $k_{\rm AB}t+a$. Labels are scaling factors used for clarity of presentation. (b) Free energy of NH$_3$ obtained from standard OPES simulations.}
\end{figure}

We are now ready to investigate the effect of water on the ammonia inversion. Thus, we repeat the calculation described above using the same setup for what concerns the path discretization. This time however, the ammonia molecule is immersed in a solution of 215 water molecules at $T$=300 K. From the trajectories thus obtained we calculate $C(t)$ and extract an inversion rate of $k_{\rm AB}^{\rm wat}\approx$6$\times$10$^{-14}$ ps$^{-1}$. This has to be compared with the rate in vacuum computed with the same force field and at the same temperature, $k_{\rm AB}^{\rm vac}\approx$3$\times$10$^{-11}$ ps$^{-1}$ [see figure~\ref{fig5.dry-VS-wet}(a)]. As to be expected, the rate is lower in solution. This reduction is in quantitative agreement with the prediction of transition state theory that gives $k_{\rm AB}^{\rm wat}/k_{\rm AB}^{\rm vac}$=$e^{-\Delta F/k_BT}\approx$2.2$\times$10$^{-3}$, where $\Delta F\approx$15 kJ/mol is the free energy difference in barrier height [see figure~\ref{fig5.dry-VS-wet}(b)].

\begin{figure}[t]\label{fig6.map}
  \centering
  \includegraphics[width=0.8\columnwidth]{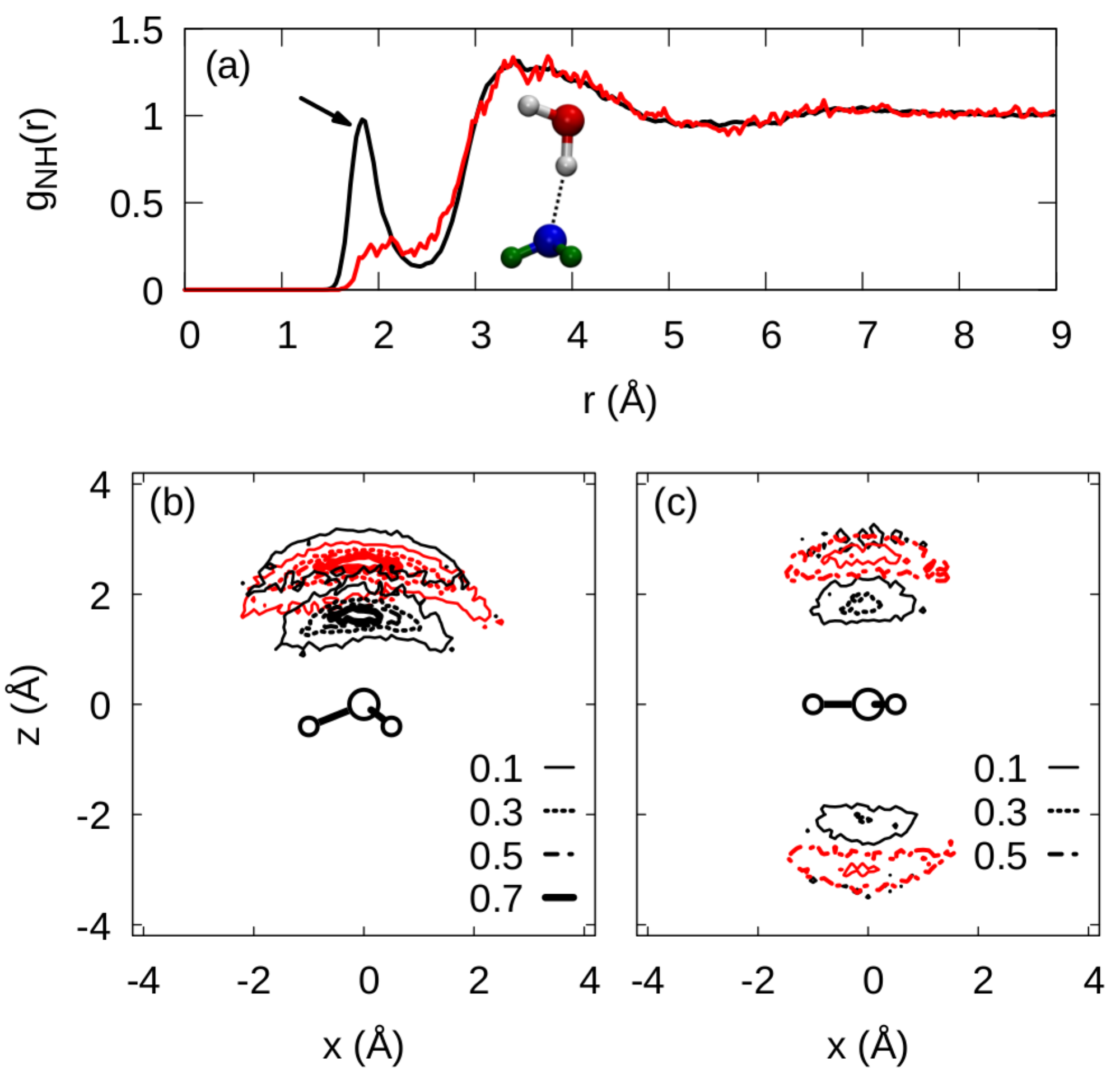}
  \caption{(a) NH pair correlation function extracted from non-reactive (black) and reactive (red) trajectories. (b), (c) Histogram of the positions of hydrogen (black) and oxygen (red) atoms of water molecules with at least one atom at a distance $<$2.5 \AA\, from nitrogen, projected onto a plane perpendicular to the hydrogen plane of NH$_3$~\cite{SM}. Histograms have been normalized such that the maximum value in panel (b) is equal to 1. Isoline values are reported in the legend. (b) Results of standard equilibrium MD simulations. (c) Results of standard MD simulations in which NH$_3$ is forced to be flat. The balls-and-sticks models show the average position of ammonia.}
\end{figure}

These positive results encourage us to study the behaviour of water during the transition. A first hint as to the role of water is given by a study of the NH$_3$-water correlations. This analysis is conducted by separating the trajectories in reactive and non-reactive. For each class of trajectories we calculate the NH pair correlation function. This is reported in figure~\ref{fig6.map}(a). In the non-reactive trajectories (black curve) there is a clear peak at $r\approx$1.8 \AA, which reflects the formation of a water-ammonia H-bond (see inset). In the reactive trajectories (red curve) this bond appears to be weakened. Further analysis shows that there is a change also in the solvation structure. While the non-reactive trajectories exhibit a non symmetric solvation shell, in the reactive ones the solvation shell is symmetric~\cite{SM}.

This behaviour can be understood by analyzing the solvation structure of the equilibrium and of the transition state configuration separately. For the equilibrium configuration, the formation of the NH$_3$-water H-bond favors one side of the solvation shell [see figure~\ref{fig6.map}(b)]. In contrast, if we artificially force NH$_3$ to be flat, as in the transition state, the solvation shell becomes symmetric and the NH$_3$-water H-bond is broken [see figure~\ref{fig6.map}(c)]. Thus, the change in solvation structure from asymmetric to symmetric lowers the transition state energy and promotes the reaction.

To conclude, we have presented a method to sample RPs via biased MD simulations in path space. The use of enhanced sampling techniques enables unconstrained exploration of RPs, making this approach more robust against problems like path trapping in metastable states. Time correlation functions can be computed via straightforward (reweighted) averages along a single MD trajectory and dynamical information such as kinetic rates are accessible at no additional cost.

In the present work we have adopted MetaD and OPES as biasing schemes, but any other enhanced sampling schemes could be applied as well. As in all biased MD simulations, prior knowledge on the mechanisms underlying the transition of interest is needed in order to build successful CVs. This is crucial to speed up convergence. It is encouraging that in the cases studied here a suitably defined end-to-end distance performed well. This represents the most natural choice. Complex systems will require a more fine tuning. However, this should not pose a major problem as one can draw from the vast literature on the subject~\cite{Valsson2016}.

Finally, we note that the proposed path approach effectively realizes parallelization of a serial problem like time evolution~\cite{Rosa-Raices2019}. This, in turn, allows a highly parallel implementation~\cite{Calhoun1996} that takes full advantage of modern massively parallel computer architectures. Given the increasing availability of massive parallel computational resources, we believe that this method will find successful applications in many fields including the study of chemical reactions, via implementation within the Car-Parrinello MD approach~\cite{Car1985}, and of the kinetics of enzymes and other biological systems.
%%%%%% BIBLIOGRAPHY %%%%%%%
\bibliographystyle{apsrev4-1}
%\bibliography{biblio}

%
\newpage
\section{Supplemental Material for ``Metadynamics of Paths''}
\section{Algorithmic details}
In the main text, we considered a molecular system whose dynamics is described by the Smoluchowski equation
\begin{equation}
\label{eq1}
m_j\nu\dot {\bf r}_j={\bf F}_j+\xi,
\end{equation}
where ${\dot {\bf r}}_j$ and $m_j$ are the velocity and mass of atom $j$, ${\bf F}_j$=$-\nabla_{{\bf r}_j}V({\bf R})$ is the force acting on it while $\nu$ is a friction coefficient and $\xi$ is a white noise term. Under this assumption, the probability of observing a discretized trajectory ${\bf R}^1\rightarrow{\bf R}^2\rightarrow\dots\rightarrow{\bf R}^N$ is given by
\begin{equation}\label{eq0}
P\propto\exp\left[-\beta V_{\rm eff}({\bf R}^1,{\bf R}^2,\dots,{\bf R}^N)\right],
\end{equation}
where $\beta$=$1/k_BT$ and the effective potential $V_{\rm eff}$ is formally equivalent to that of an open polymer of $N$ beads ${\bf R}^n$, corresponding to the configurations visited along the trajectory at times $t_n=(n-1)\Delta t$ (see equation (5) of the main text).

In order to sample trajectories distributed according to equation~\eqref{eq0}, we adopted a Hamiltonian approach. That is, we introduced auxiliary momenta \{${\bf p}_j^n$\} and masses \{${\rm M}_j$\} and performed molecular dynamics (MD) simulations solving Hamilton's equations
\begin{eqnarray}\label{eq1.Hp}
\dot {\bf p}^{n}_j&=&-\nabla_{{\bf r}^{n}_j} V_{\rm eff}\\
\dot {\bf r}^{n}_j&=&\frac{{\bf p}^{n}_j}{{\rm M}_j} \label{eq1.Hr}
\end{eqnarray}
coupled to a thermostat.

Equation~\eqref{eq1.Hp} involves terms that depend on the second derivatives of the potential $V({\bf R})$. These are evaluated using the following symmetric finite difference formula~\cite{Putrino2000,Kapil2016}
\begin{flalign}\label{eq}
\sum_{k=1}^{M}\sum_{\alpha=x,y,z}\frac{\partial F_{k,\alpha}^n}{\partial r_{j,\beta}^n}\eta_{k,\alpha}^n\approx&&\nonumber
\end{flalign}
\begin{flalign}
\frac{1}{2\varepsilon}\left[F_{j,\beta}^n(r_{k,\alpha}^n+\varepsilon \eta_{k,\alpha}^n)-F_{j,\beta}^n(r_{k,\alpha}^n-\varepsilon \eta_{k,\alpha}^n)\right],
\end{flalign}
where $\eta_{k,\alpha}^n=r_{k,\alpha}^{n+1}-r_{k,\alpha}^n-\frac{\Delta t}{m_k\nu} F_{k,\alpha}^n$ and $\varepsilon$ is a number small enough to guarantee energy conservation in microcanonical simulations. Equation~\eqref{eq} amounts to a modest but necessary increase in computational cost as it avoids direct implementation of the Hessian $\partial^2 V({\bf R})/\partial r_{j,\alpha}\partial r_{k,\beta}$.
\section{Computation of dynamical quantities}\label{kkk}
Each time step of the fictitious dynamics~\eqref{eq1.Hp}, \eqref{eq1.Hr} generates a new polymer configuration $\{{\bf R}^1,{\bf R}^2,\dots,{{\bf R}^N}\}$ corresponding to a full discretized trajectory ${\bf R}^1\rightarrow{\bf R}^2\rightarrow\dots\rightarrow{\bf R}^N$ of the original system. The average value of an observable $O({\bf R})$ at time $t_n$ is computed as
\begin{equation}
\langle O^n\rangle=\frac1 Z\sum_{i=1}^ZO_i^n,
\end{equation}
where $Z$ is the total number of polymer configurations sampled and $O^n=O({\bf R}^n)$ is the value of the observable evaluated in the $n$-th bead. Similarly, time correlation functions
\begin{equation}
C_{OO}(t_{n'}-t_n)=\langle O^nO^{n'}\rangle=\frac1 Z\sum_{i=1}^ZO_i^nO_i^{n'}
\end{equation}
are computed as average values of the product between observables evaluated in different beads of the fictitious polymer. In Metadynamics (MetaD) simulations, the correct statistics is obtained computing reweighted averages~\cite{Bonomi2009,Branduardi2012,tiwary2015,Mones2016,Marinova2019,invernizzi2020,giberti2020}.

The phenomenological rate constant $k_{\rm AB}$ can be extracted from simulations computing the time correlation function $C(t)$ (equation (6) of the main text). In absence of intermediate states and after a short transient time $\tau_{\rm trans}$, $C(t)$ enters a linear regime and the rate is given by its slope
\begin{equation}
\label{eqK}
k_{\rm AB}=\frac{{\rm d}C(t)}{{\rm d}t}.
\end{equation}
Equation~\eqref{eqK} is valid at times $\tau_{\rm trans}\ll t \ll k_{\rm AB}^{-1}$, larger than the transient, but short compared to the characteristic relaxation time of the system~\cite{Chandler1987}.
\begin{figure}[tb]\label{100-VS-200}
  \centering
  \includegraphics[width=0.9\columnwidth]{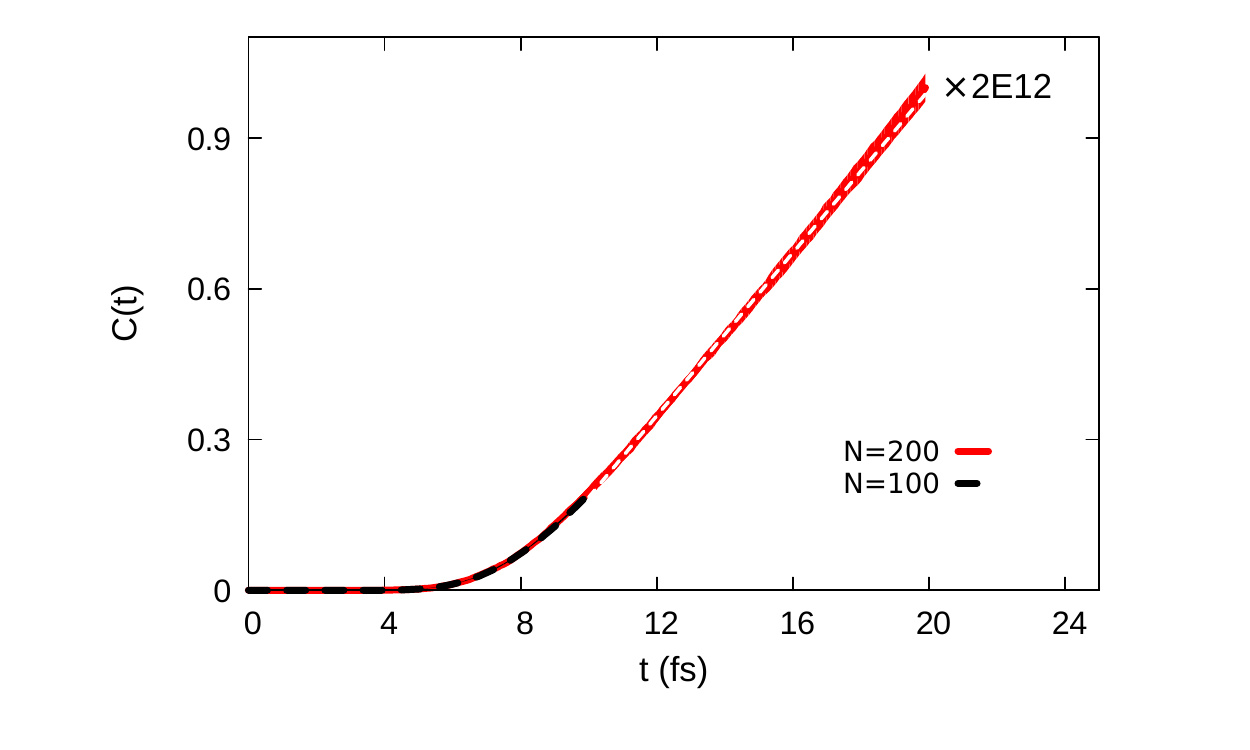}
  \caption{Umbrella inversion of NH$_3$ in vacuum. The time correlation function obtained at $T$=300 K using polymer models of two different sizes $N$=100 and 200. The dashed white line is a linear fit to $k_{\rm AB}t+a$ of the data at $N$=200. The model and the simulation protocol are discussed in section~\ref{water}. The label is a scaling factor applied to both curves.}
\end{figure}
\section{Choice of the polymer model parameters}
The parameter defining the polymer model discretization is the ratio $\frac {\nu} {\Delta t}$ between the damping coefficient and the time step. In practice, in our simulations we fixed $\nu$ and we selected the smallest time step $\Delta t$ for which we could converge the results. We ensured that the adopted time step is compatible with the dynamics of the original system. This was done checking the stability of the trajectories in standard Langevin simulations performed in the overdamped limit.

In order to extract the kinetic rate, the size $N$ of the polymer model must be large enough to observe the onset of the linear regime of $C(t)$. This requirement sets the minimum value $N_{\rm min}$ needed in simulations. The rate $k_{\rm AB}$=${\rm d}C(t)/{\rm d}t$ should be independent of the polymer size, as long as $N\gtrsim N_{\rm min}$. To check this, we have performed simulations considering polymer models of increasing size. In figure~\ref{100-VS-200} we report results for the case of the umbrella inversion of NH$_3$ in vacuum obtained using two different values of $N$=100 and 200. As expected, the curves show very good agreement. A linear fit yielded values of $k_{\rm AB}\approx$(3.2$\pm$0.1)$\times$10$^{-11}$ ps$^{-1}$ and $k_{\rm AB}\approx$(3.9$\pm$0.1)$\times$10$^{-11}$ ps$^{-1}$, respectively for $N$=100 and 200. The latter estimate represents the converged result. Given the illustrative purposes of the application, we eventually decided to perform most simulations using a polymer of size $N$=100. This reduced the computational cost while still yielding reasonably converged values of the kinetic rates.
\section{2D double-well potential}\label{2D}
\subsection{Model and simulation setup}\label{2D.setup}
In the first application, we considered the dynamics of a particle in the two dimensional potential~\cite{Dellago1998a}
\begin{equation}
U(x,y)= 2 + \frac 4 3 x^4 - 2y^2 + y^4 + \frac {10} {3} x^2(y^2-1).
\end{equation}
Simulations were performed using the following set of parameters for the trajectory discretization: $m$=1, $\Delta t$=0.15, $\nu$=1. In all simulations, we considered a polymer of size $N$=200 inside a square cell of side $L$=10 centered at the origin. The exact analytical expression of the force governing the dynamics of the polymer has been hard coded in LAMMPS~\cite{Plimpton1995}. The equations of motion in trajectory space were solved adopting a standard velocity-Verlet integrator with a time step of $\Delta t_{\rm MD}$=0.01 and auxiliary masses set to $M$=1. Temperature was controlled via a Nos\'e-Hoover chains thermostat~\cite{Martina1992}. The PLUMED~\cite{Tribello2014} enhanced sampling library was used to introduce harmonic restraints and to perform well-tempered MetaD~\cite{Barducci2008,Dama2014} (WT-MetaD) as well as OPES~\cite{invernizzi2020} simulations. Here and in the main text, all quantities are reported as obtained from simulations with the above set of adimensional parameters.
\subsection{MetaD in trajectory space and dynamical TPS}\label{2D.meta}
In the first set of simulations, we investigated the ability of our MetaD approach to sample different reactive pathways (RPs) in the same run and we compared results with those obtained adopting the dynamical transition path sampling (TPS) algorithm of Ref.~\onlinecite{Dellago1998a}. TPS simulations were performed at temperature $k_BT$=0.05, applying harmonic constraints (spring constant $K$=100) to the two distances $R_{1,{\rm A}}$=$\lvert {\bf R}^1-{\bf R}_{\rm A} \rvert$ and $R_{N,{\rm B}}$=$\lvert {\bf R}^N-{\bf R}_{\rm B} \rvert$. Here, ${\bf R}^{1,N}$ are the positions of the first and last bead of the polymer, while ${\bf R}_{\rm A,B}\approx(\mp1,0)$ mark the two minima of the potential. With this choice of parameters, $R_{1,{\rm A}}$ and $R_{N,{\rm B}}$ were bound to values $\leq$0.1. WT-MetaD simulations were performed at the same temperature, applying same harmonic constraint only to $R_{1,{\rm A}}$. The polymer end-to-end distance $d_{\rm e2e}$=$\lvert {\bf R}^1-{\bf R}^N\rvert$ was used as collective variable (CV). The bias potential was built using a bias factor $\gamma$=20, depositing Gaussian kernels (height=1, $\sigma$=0.1) every 2500 MD steps. In both cases, we performed simulations of 5$\times$10$^{7}$ MD steps, sampling configurations every 2500 steps. In the case of TPS, all polymer configurations sampled were used for subsequent analysis. In the case of MetaD, we considered only the second half of the trajectory (well within the asymptotic regime of WT-MetaD) and defined successful RPs those for which $R_{1,{\rm A}}$$\leq$0.1 and $R_{N,{\rm B}}$$\leq$0.1 at the same time. These were used for the analysis. Each experiment was repeated 5 times. Figures 2(c) and 2(d) of the main text report the average value of the probability distribution extracted from the 5 independent runs.
\subsection{Computation of kinetic rates}\label{2D.rates}
\begin{figure}[tb]\label{fig.trace2D}
  \centering
  \includegraphics[width=0.8\columnwidth]{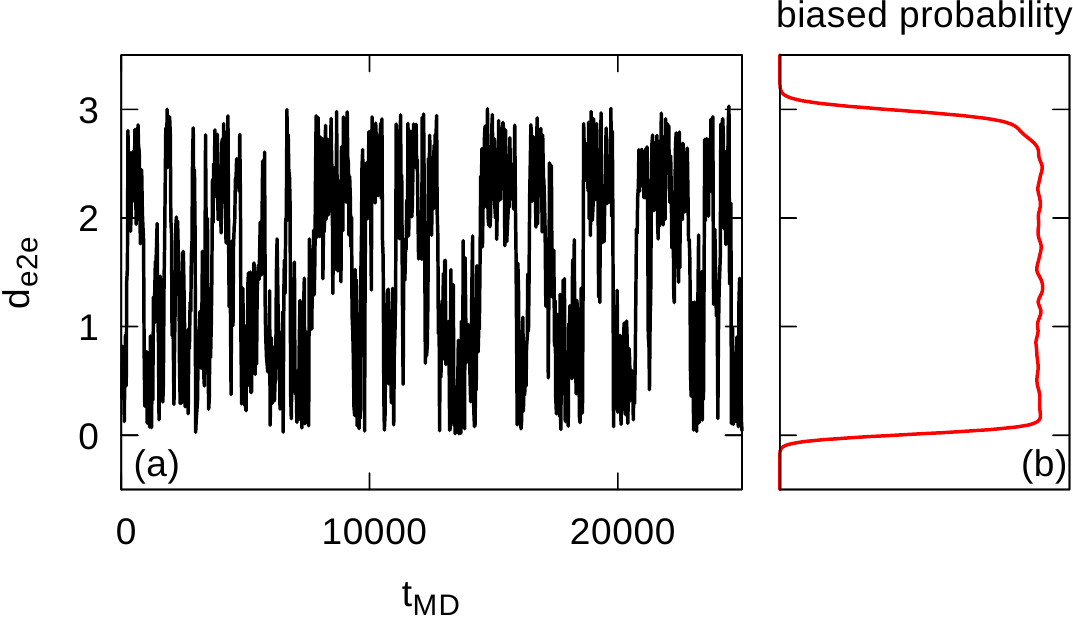}
  \caption{(a) Time evolution of $d_{\rm e2e}$ at $k_BT$=0.125. (b) The corresponding biased probability distribution.}
\end{figure}

In the second set of simulations (not discussed in the main text), we adopted OPES~\cite{invernizzi2020} to compute the phenomenological kinetic rate. Simulations were performed at temperature $k_BT$=0.125, 0.2, 0.3, 0.4 and 0.5, without applying any constraint to the polymer. We set the biasfactor to $\gamma$=$\infty$ and deposited Gaussian kernels every 2500 MD steps. The standard deviation of the kernels was set equal to $\sigma$=0.1 at $k_BT$=0.125 and to $\sigma$=0.2 at all other temperatures. In OPES, the height of the kernels is automatically adjusted during runtime. The last input parameter is an estimate of the free energy barrier to be overcome, which we set equal to $\Delta F$=1.5 for $k_BT$=0.125, 0.2, 0.3 and to $\Delta F$=1.2 for $k_BT$=0.4, 0.5. These values were estimated from preliminary WT-MetaD simulations. With this setup, we effectively targeted a uniform probability distribution of $d_{\rm e2e}$ in the interval 0$\lesssim d_{\rm e2e}\lesssim$3 (see figure~\ref{fig.trace2D}). 
\begin{figure}[tb]\label{fig.converge2D}
  \centering
  \includegraphics[width=0.8\columnwidth]{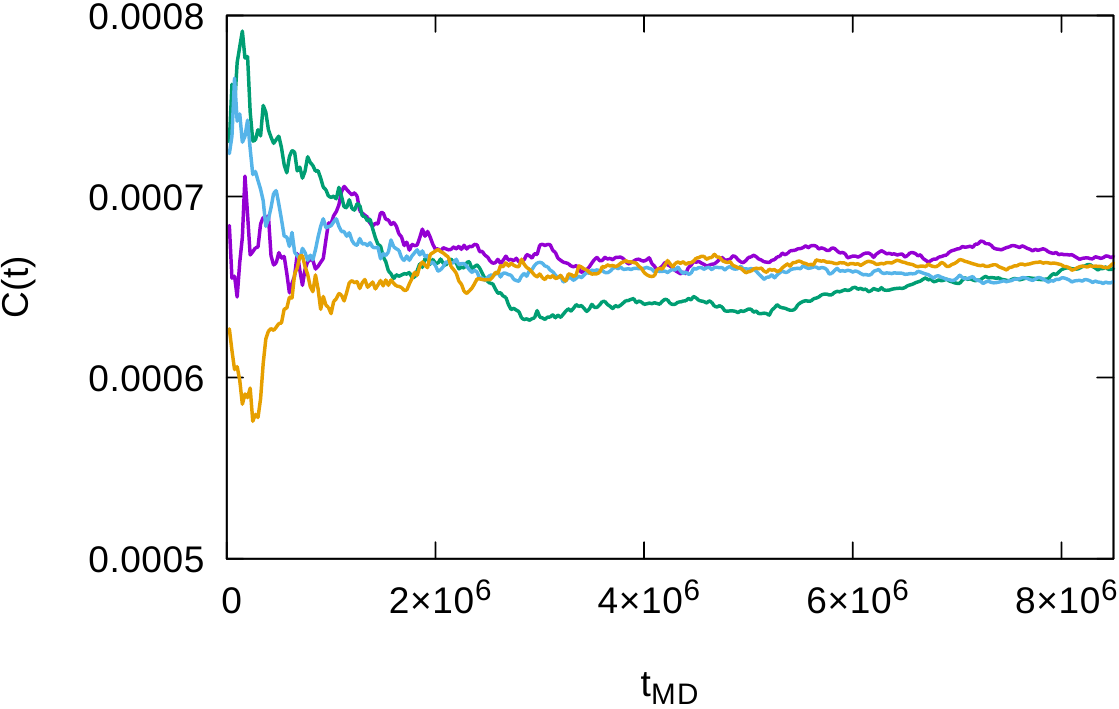}
  \caption{Time evolution of the time correlation function $C(t)$ at a lag time $t$=20, obtained from the reweighting of 4 independent simulations at $k_BT$=0.125.}
\end{figure}

\begin{figure}[tb]\label{fig3.FES2D}
  \centering
  \includegraphics[width=0.8\columnwidth]{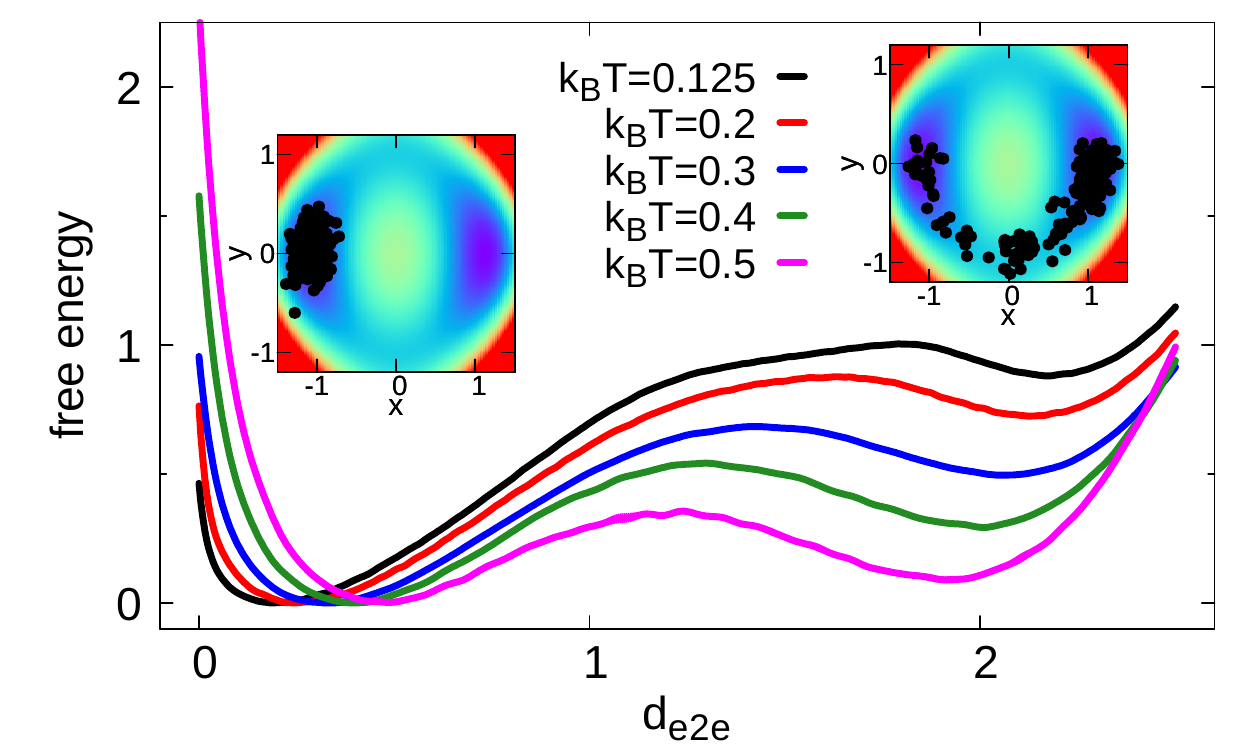}
  \caption{Free energy curves as a function of the end-to-end distance. Insets show representative polymer configurations at $d_{\rm e2e}\approx0$ and $d_{\rm e2e}\approx2$.}
\end{figure}

\begin{figure}[tb]\label{fig4.2Drate}
  \centering
  \includegraphics[width=0.8\columnwidth]{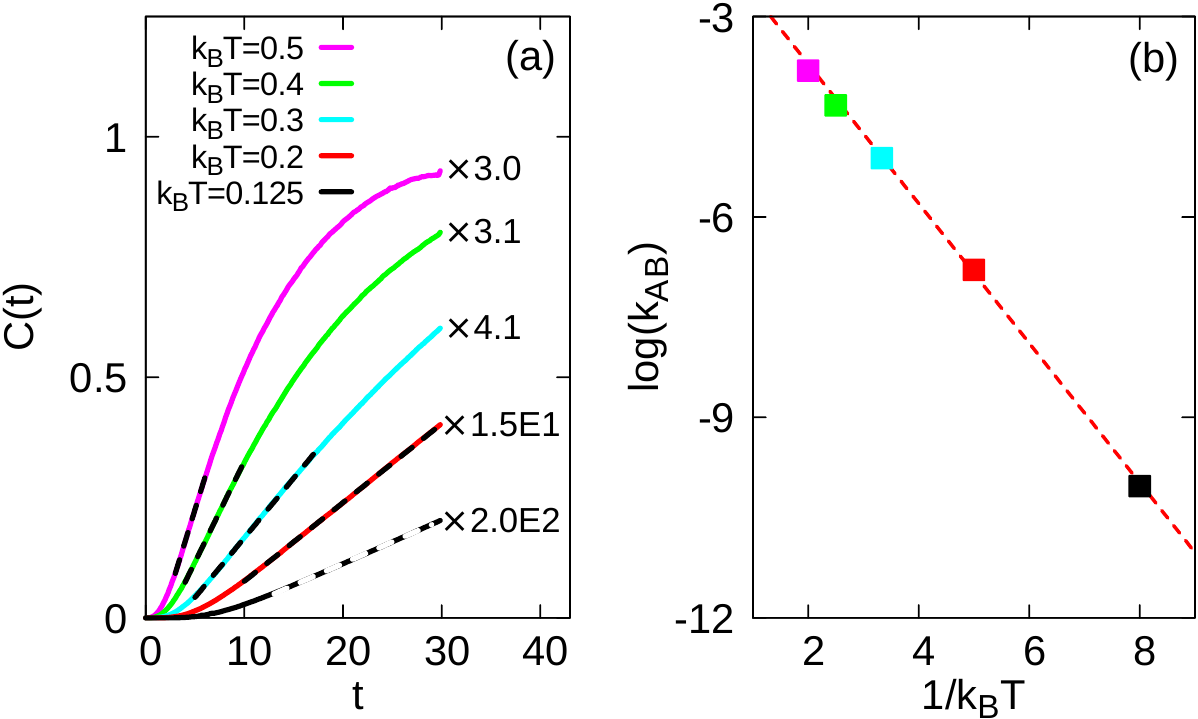}
  \caption{(a) The time correlation function, $C(t)$, computed at different temperatures. Dashed lines are fit to $k_{\rm AB}t+a$. Labels are scaling factors used for clarity of presentation. (b) Arrhenius plot of the rate constants. The red dashed line is a fit to $-\Delta E_{\rm fit}/k_BT+b$.}
\end{figure}

At each temperature, we performed a simulation of 5$\times$10$^{8}$ MD steps. The time correlation function $C(t)$ was computed defining the characteristic functions as $I_{\rm A,B}({\bf R}^n)$=1 if $\lvert {\bf R}^n-{\bf R}_{\rm A,B}\rvert<$0.7 and zero otherwise. Time averages were computed using the reweighting scheme suggested in Ref.~\onlinecite{invernizzi2020}, neglecting the initial 5$\times$10$^{6}$ steps. Each experiment was repeated 4 times. Figure~\ref{fig.converge2D} shows the time evolution of the correlation function at one lag time. Figures~\ref{fig3.FES2D} and \ref{fig4.2Drate}(a), discussed below, present average values and standard deviations estimated from the 4 independent runs.

Figure~\ref{fig3.FES2D} shows the free energy curves of the fictitious polymer as a function of the end-to-end distance. At all temperatures considered, we observe a global minimum near $d_{\rm e2e}\approx0$, whose population corresponds to trajectories that never leave the left basin (see left inset). A secondary minimum is found at $d_{\rm e2e}\approx2$, which is the distance between the two minima of the potential. Accordingly, the corresponding polymer configurations represent RPs that successfully reach B after having crossed one of the two equivalent saddles (see right inset). Increasing the temperature lowers the free energy barrier, which reflects the thermally enhanced probability for the particle to cross from A to B. At low temperatures, configurations corresponding to intermediate values of $d_{\rm e2e}\approx1$ are ``failed attempts'' of the polymer that stretches towards the right basin without reaching it. On the other hand, for $k_BT\geq$0.4 we observe contributions also from configurations recrossing from B to A.

In figure~\ref{fig4.2Drate}(a) we show the correlation function obtained at various temperatures, displaying the expected short transient, followed by linear growth. Note that for $k_BT\geq$0.4, recrossings of the polymer from B to A are observed at times $t\ge$10. This explains the deviation of $C(t)$ from the linear trend at times $t\ge$10, where the conditions for equation~\eqref{eqK} to hold cease to apply~\cite{Chandler1987}. Nevertheless, we were able to extract the phenomenological rate constant $k_{\rm AB}$ in the whole range of temperatures by considering the initial linear regime. These are reported in the Arrhenius plot of figure \ref{fig4.2Drate}(b), showing the expected linear trend. A linear fit of the data yielded an activation energy of $\Delta E_{\rm fit}=1.05\pm0.02$, in agreement with the exact value of the potential barrier $\Delta E\approx1.08$.
\section{NH$_3$ in vacuum}\label{vacuum}
\subsection{Model and simulation setup}\label{NH3vac.meta}
\begin{figure}[tb]\label{fig.pes}
  \centering
  \includegraphics[width=0.8\columnwidth]{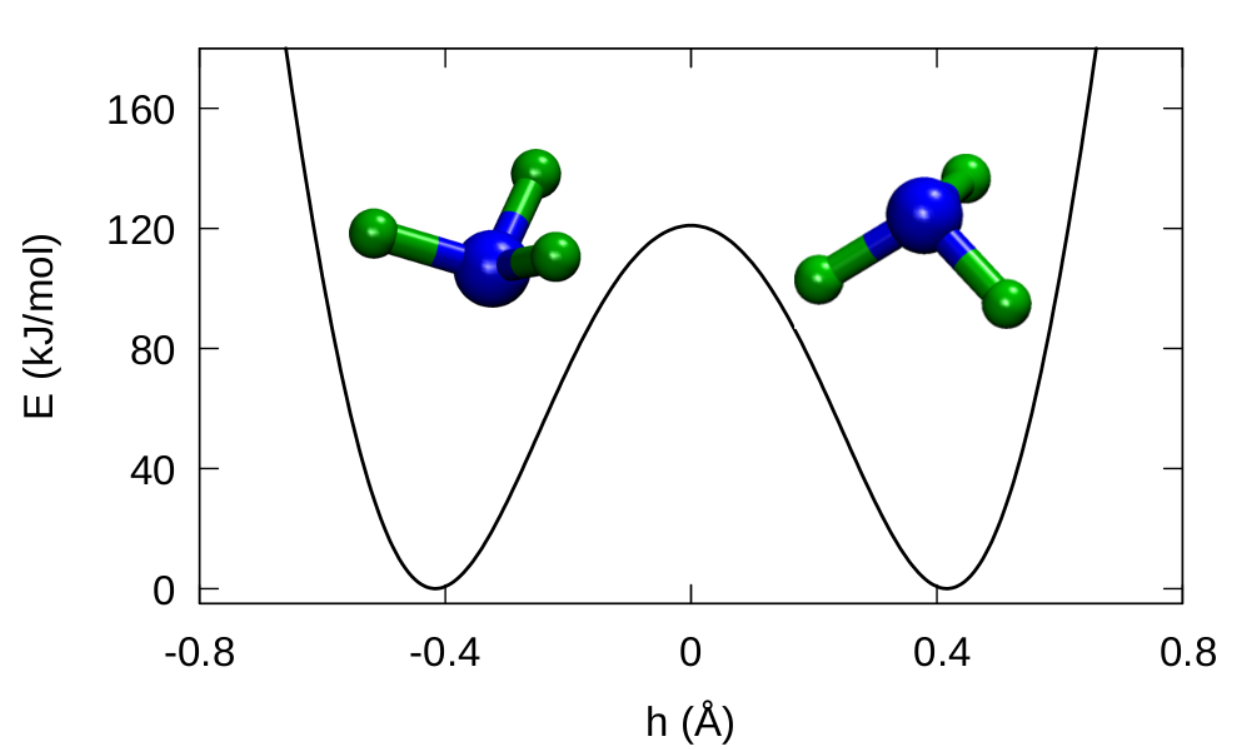}
  \caption{Potential energy of NH$_3$ as a function of its oriented heigth $h$.}
\end{figure}

In the second application, we considered the umbrella inversion of ammonia in vacuum. We considered a single NH$_3$ molecule in open boundary conditions. Intra-molecular interactions were described using the ReaxFF force field of Ref.~\onlinecite{Weismiller2010}, neglecting electrostatics. Figure~\ref{fig.pes} shows the potential energy profile obtained via a sequence of geometry optimizations at fixed values of the oriented height $h$ of the NH$_3$ tetrahedron. The two symmetric minimum energy confingurations at $h$$\approx\pm$0.4 \AA\, are separated by a barrier of $\approx$120 kJ/mol.

Simulations in trajectory space were performed using a polymer of size $N$=100 beads and the following set of parameters for the trajectory discretization: $\Delta t$=0.1 fs, $\nu$=0.14 fs$^{-1}$. The auxiliary masses $M_j$ were set equal to those of the corresponding atom in the bead. The equations of motion have been implemented in LAMMPS~\cite{Plimpton1995} and solved adopting a velocity-Verlet algorithm with an MD integration step of $\Delta t_{\rm MD}$=0.25 fs. Terms in the forces containing second derivatives of the potential energy were estimated adopting the finite difference expression~\eqref{eq}. Temperature was controlled via a Nos\`e-Hoover chains thermostat~\cite{Martina1992}.
\begin{figure}[tb]\label{fig2.trace-e2e}
  \centering
  \includegraphics[width=0.8\columnwidth]{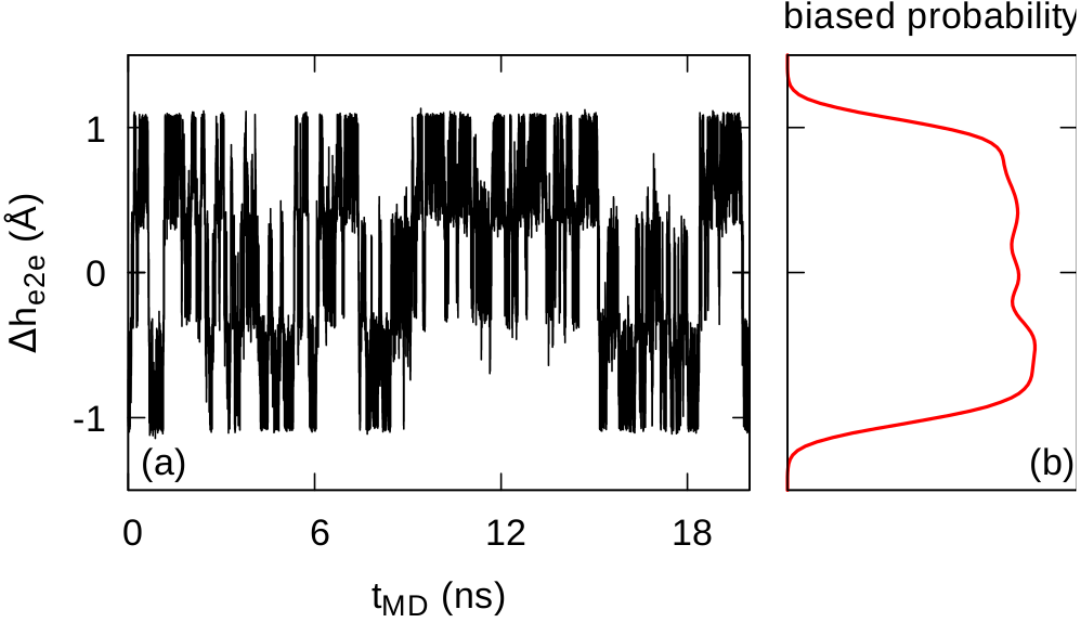}
  \caption{(a) Time evolution of $\Delta h_{\rm e2e}$ at $T$=500 K. (b) The corresponding biased probability distribution.}
\end{figure}
\begin{figure}[tb]\label{fig.convergeNH3}
  \centering
  \includegraphics[width=0.8\columnwidth]{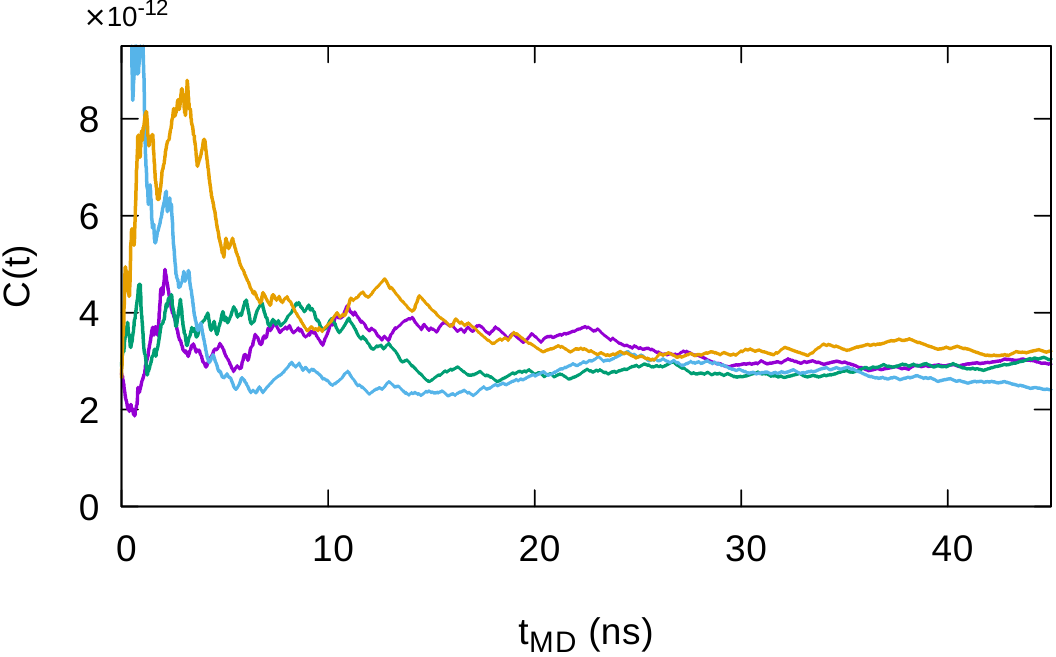}
  \caption{Time evolution of the correlation function $C(t)$ at a lag time $t$=8 fs, obtained from the reweighting of 4 independent simulations at $T$=500 K.}
\end{figure}
\begin{figure}[tb]\label{figfes}
  \centering
  \includegraphics[width=0.8\columnwidth]{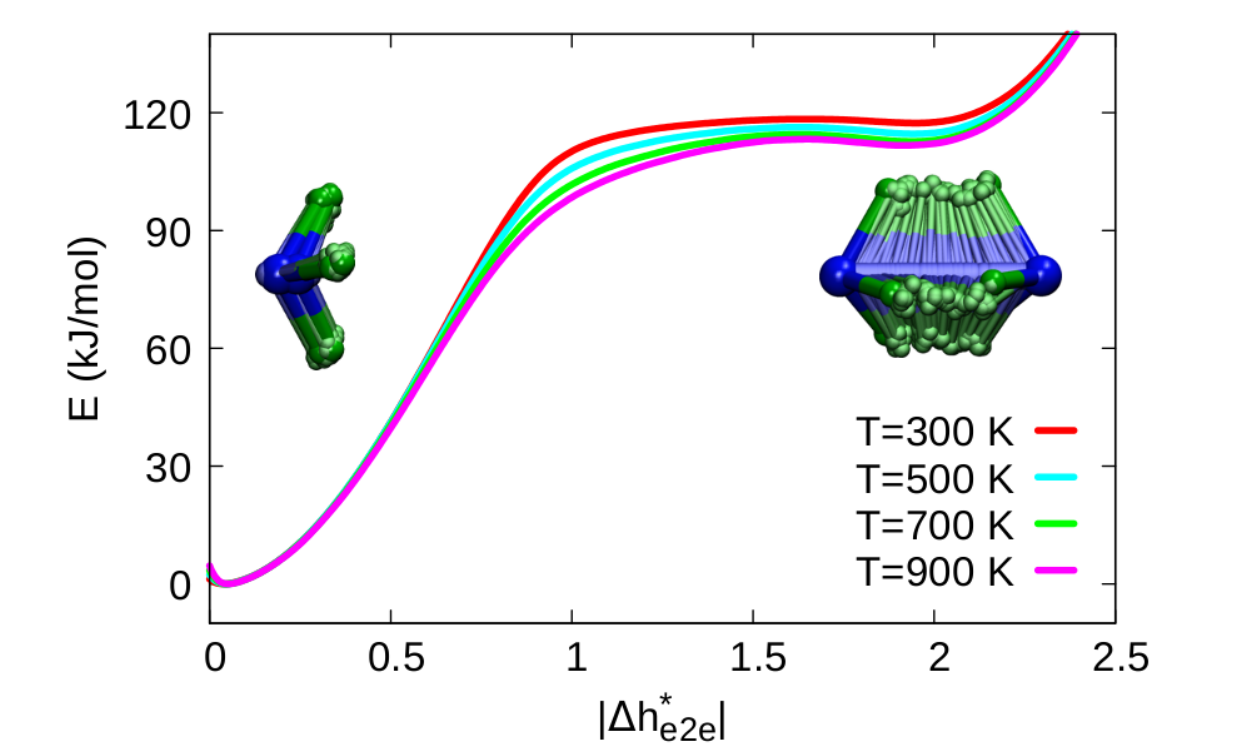}
  \caption{Free energy curves as a function of the generalized end-to-end distance $\lvert\Delta h_{\rm e2e}\rvert$. Insets show representative polymer configurations at $\lvert\Delta h_{\rm e2e}\rvert$$\approx$0 and $\lvert\Delta h_{\rm e2e}\rvert$$\approx$0.8 \AA. Replicas have been suitably aligned for clarity of presentation.}
\end{figure}
\subsection{Computation of kinetic rates}
Biased MD simulations in trajectory space were performed using OPES~\cite{invernizzi2020}. We considered temperatures of $T$=300, 500, 700, 900 K. We chose as CV the generalized polymer end-to-end distance $\Delta h_{\rm e2e}=(h^N-h^1)$, equal to the difference between the oriented height of the last and of the first ammonia replica. We used a bias factor $\gamma$=$\infty$ and deposited Gaussian kernels every 500 MD steps. The standard deviation of the kernels was set equal to $\sigma$=0.0032, 0.0043, 0.0051, 0.0059 \AA, respectively for $T$=300, 500, 700, 900 K. The barrier parameter was set equal to $\Delta F$=140 kJ/mol. We did not apply any restraint to the polymer. With this setup, we effectively targeted a uniform probability distribution of $\Delta h_{\rm e2e}$ in the interval -1 \AA$\lesssim \Delta h_{\rm e2e}\lesssim$1 \AA\, (see figure~\ref{fig2.trace-e2e}).

At each temperature, we ran a simulation of $\approx$3$\times$10$^8$ MD steps. The time correlation function $C(t)$ was computed defining the characteristic function of the two basins $I_{\rm A,B}(h)$=1 if $\lvert h\rvert>0.2$ \AA\, and zero otherwise. Time averages were computed using the reweighting scheme suggested in Ref.~\onlinecite{invernizzi2020}, skipping the initial $\approx$10$^7$ MD steps. Each experiment was repeated 4 times. Figure~\ref{fig.convergeNH3} shows the time evolution of the correlation function evaluated at one lag time. Figures 3(a) of the main text and figure~\ref{figfes}, discussed below, present average values and standard deviations estimated from the 4 independent runs.

Figure~\ref{figfes} reports the free energy curves of the fictitious polymer obtained at $T$= 300, 500, 700, 900 K. At all temperatures considered, we observe a global minimum near $\lvert\Delta h_{\rm e2e}\rvert$$\approx$0, whose population corresponds to non-reactive trajectories where all replicas share the same orientation (see left inset). That is followed by a plateau and by a secondary shallow minimum at $\lvert\Delta h_{\rm e2e}\rvert$$\approx$0.8 \AA, which is the difference between the values of $h$ in the two symmetric equilibrium states. Accordingly, the corresponding polymer configurations represent RPs where ammonia flips between the $h$=$\pm$0.4 \AA\, states (see right inset).
\section{NH$_3$ in water}\label{water}
\subsection{Model and simulation setup}\label{NH3wat.meta}
In the third application, we investigated the umbrella inversion of NH$_3$ in water at $T$=300 K. We used a cubic box of size $L\approx$18.6 \AA\, containing 215 water molecules (density $\approx$1 g/cm$^3$) and one ammonia molecule, and we enforced periodic boundary conditions. Intra-molecular interactions of NH$_3$ were described using the following force field:
\begin{equation}
V_{\rm NH_3}=\sum_{i=1}^3\left(D\left[1-e^{-\alpha(r_i-r_0)^2}\right]+K(\theta_i-\theta_0)^2\right),
\end{equation}
where $r_i$ and $\theta_i$ indicate the three N-H distances and H-N-H angles. We used the following set of parameters: $D$=101.905 kcal/mol, $\alpha$=2.347 \AA$^{-1}$, $r_0$=1.0124 \AA, $K$=103.045 kcal/mol\AA$^2$, $\theta_0$=106.67$^\circ$. Fixed partial charges of $q_{\rm H}$=0.342$e$ and $q_{\rm N}$=-1.026$e$ were assigned to the hydrogen and nitrogen atoms. Water was described using the tip3p model~\cite{jorgensen1983} with partial charges of $q_{\rm H}$=0.415$e$ and $q_{\rm O}$=-0.83$e$. Inter-molecular van der Waals interactions of the form
\begin{equation}
V(r)=4\varepsilon\left[\left(\frac{\sigma}{r}\right)^{12}-\left(\frac{\sigma}{r}\right)^6\right]
\end{equation}
were computed within a cutoff distance of $r_{\rm cut}$=9 \AA. We used the following set of parameters: $\varepsilon$=0.102, 0.21, 0.1463557 kcal/mol and $\sigma$=3.188, 3.36, 3.2728703 \AA, respectively for O-O, N-N and N-O pairs, and zero otherwise. Long range electrostatic interactions were computed using the particle-particle particle-mesh solver~\cite{Hockney} as implemented in LAMMPS~\cite{Plimpton1995}, with an accuracy of $10^{-4}$ kcal/mol\AA. In all simulations, temperature was controlled using a Langevin thermostat with time constant $\tau$=1 ps. The equations of motions were integrated using a velocity-Verlet algorithm and a time step of $\Delta t_{\rm MD}$=0.5 fs. For comparison, we also perform simulations without water, using the same setup.
\begin{figure}[tb]\label{frame}
  \centering
  \includegraphics[width=0.7\columnwidth]{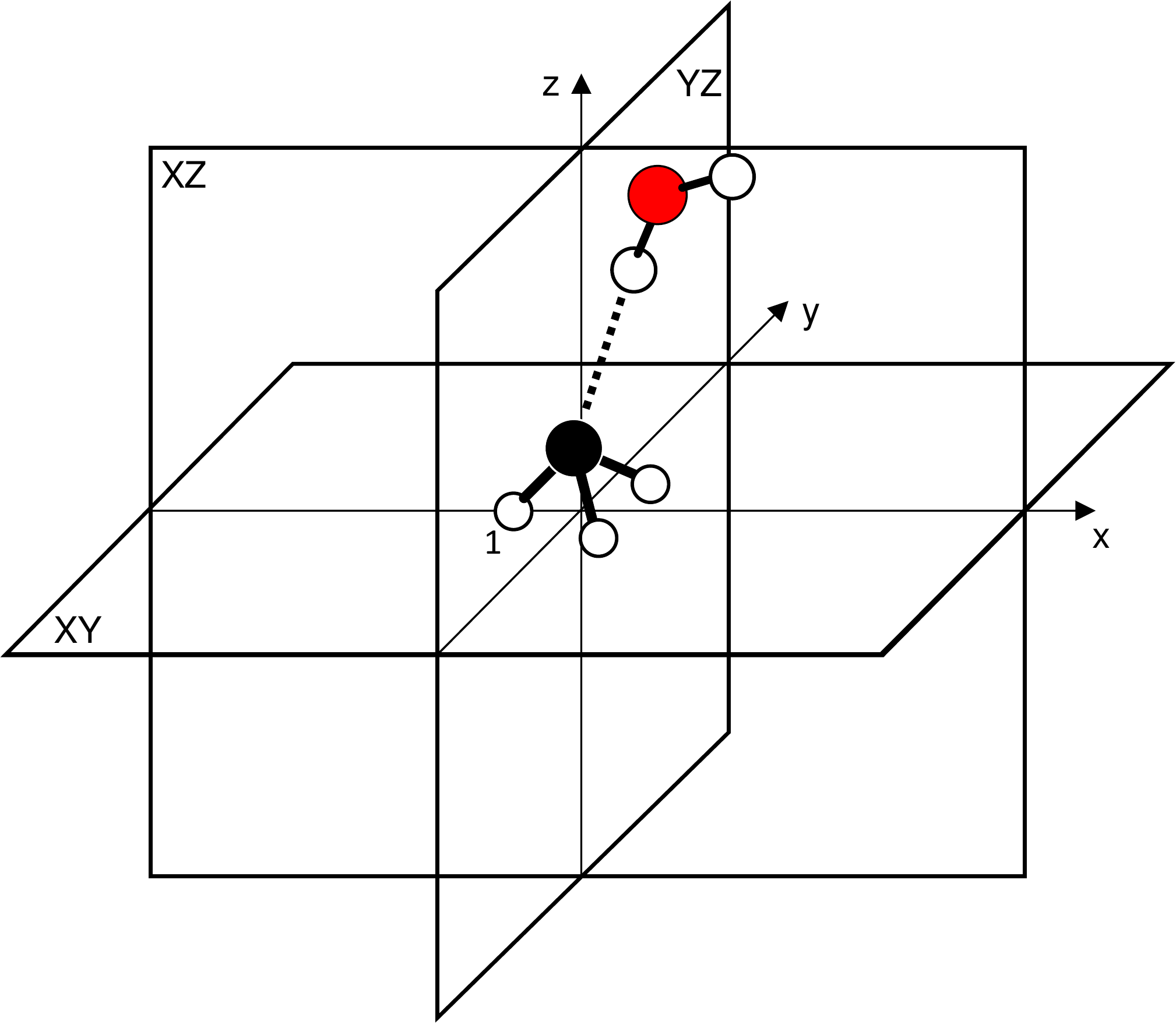}
  \caption{The Cartesian reference frame used for the analysis of the solvation shell of NH$_3$. The hydrogen plane of ammonia defines the ($x$,$y$) plane, the height of the tetrahedron defines the $z$-axis and one hydrogen atom (marked 1) is placed along the $x$-axis.}
  \end{figure}
\begin{figure}[tb]\label{mappe}
  \centering
  \includegraphics[width=0.8\columnwidth]{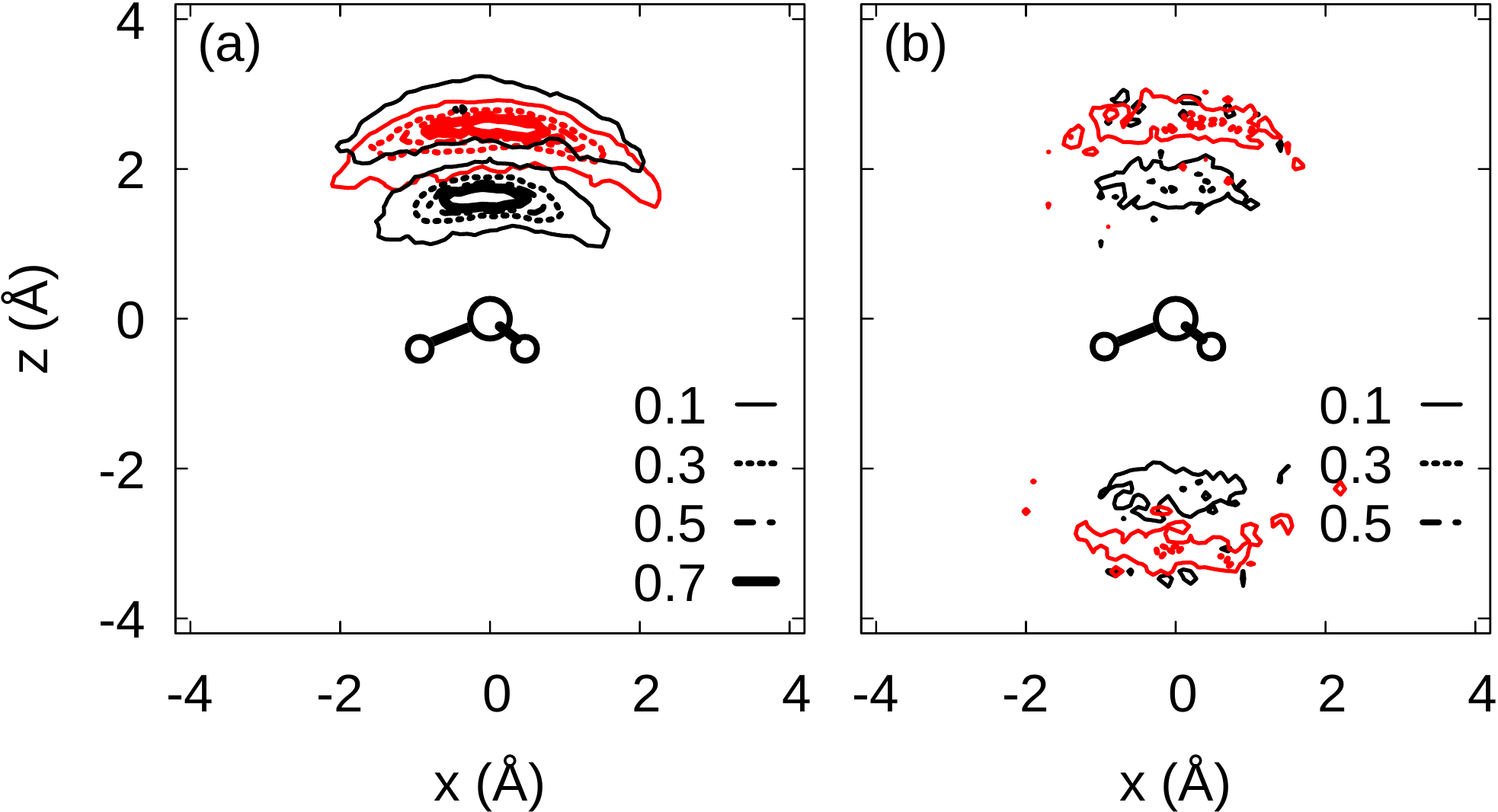}
  \caption{Histogram of the positions of hydrogen (black) and oxygen (red) atoms of water molecules with at least one atom at a distance $<2.5$ \AA\, from nitrogen, projected onto the ($x$, $z$) plane (see figure~\ref{frame}). Histograms have been normalized such that the maximum value in panel (a) is equal to 1. Isoline values are reported in the legend. (a) Results obtained considering configurations extracted from non-reactive trajectories. (b) Results obtained considering the starting configurations of reactive trajectories. The balls-and-sticks models show the average position of ammonia.}
\end{figure}
\subsection{Free energy of NH$_3$}
In order to investigate the effect of water on the free energy barrier for the umbrella transformation, we perfomed standard OPES simulations in configuration space with and without water. We used the oriented height $h$ as CV, we set the biasfactor to $\gamma$=30, the barrier parameter to $\Delta F$=95 kJ/mol and we deposited kernels ($\sigma$=0.028 \AA) every 2000 steps. We performed 4 independent simulations of 2$\times$10$^6$ steps and computed the free energy curves using the reweighting scheme of Ref.~\onlinecite{invernizzi2020}. Figure 4(b) of the main text reports the corresponding average values and standard deviations.
\subsection{Computation of kinetic rates}
Simulations in trajectory space were performed adopting a polymer of size $N$=100 beads and the following parameters for the trajectory discretization: $\Delta t$=0.1 fs and $\nu$=0.14 fs$^{-1}$. The fictitious masses were set equal to those of the corresponding atom in the polymer bead. Biased simulations were performed using OPES with the generalized end-to-end distance $\Delta h_{\rm e2e}$=$(h^N-h^1)$ as CV. We set the biasfactor to $\gamma$=$\infty$, the barrier parameter to $\Delta F$=120 kJ/mol and we deposited kernels ($\sigma$=0.035 \AA) every 2000 steps. We performed 4 independent simulations of 3$\times$10$^7$ steps both with and without water. The time correlation function $C(t)$ was computed defining the characteristic function of the two basins $I_{\rm A,B}(h)$=1 if $\lvert h\rvert>0.2$ \AA\, and zero otherwise. Time averages were computed using the reweighting scheme suggested in Ref.~\onlinecite{invernizzi2020}, skipping the initial 4$\times$10$^6$ MD steps. Figure 4(a) of the main text presents average values and standard deviations estimated from the 4 independent runs.
\subsection{Analysis of the solvation shell of NH$_3$}
In order to analyze the solvation shell of NH$_3$, we selected all water molecules with at least one atom at a distance $<$2.5 \AA\, from nitrogen and we projected their positions onto the ($x$, $z$) plane of the Cartesian frame of reference defined in figure~\ref{frame}.

In figure~\ref{mappe}(a) we report the results of this analysis obtained considering configurations extracted from non-reactive trajectories, showing an asymmetric solvation shell due to the presence of a NH$_3$-water H-bond. In figure~\ref{mappe}(b) we report the results obtained considering the starting configurations of reactive trajectories, where the NH$_3$-water H-bond is already broken and the solvation shell is symmetric.
%%%%%% END DOCUMENT %%%%%%

\begin{thebibliography}{40}
\bibitem{Laio2002} A. Laio and M. Parrinello, PNAS {\bf 99}, 12562 (2002).
\bibitem{Tiwary2013} P. Tiwary and M. Parrinello, Phys. Rev. Lett. {\bf 111}, 230602 (2013).
\bibitem{Wu2014} H. Wu, A. S. J. S. Mey, E. Rosta, and F. No\`e, J. Chem. Phys. {\bf 141}, 214106 (2014).
\bibitem{Rosta2015} E. Rosta and G. Hummer, J. Chem. Theory Comput. {\bf 11}, 276 (2015).
\bibitem{Wu2016} H. Wu, F. Paul, C. Wehmeyer, and F. No\`e, PNAS {\bf 113}, E3221 (2016).
\bibitem{Donati2017} L. Donati, C. Hartmann, and B. G. Keller, J. Chem. Phys. {\bf 146}, 244112 (2017).
\bibitem{Donati2018} L. Donati and B. G. Keller, J. Chem. Phys. {\bf 149}, 072335 (2018).
\bibitem{Pratt1986} L. R. Pratt, J. Chem. Phys. {\bf 85}, 5045 (1986).
\bibitem{Elber1987} R. Elber and M. Karplus, Chem. Phys. Lett. {\bf 139}, 375 (1987).
\bibitem{Olender1996} R. Olender and R. Elber, J. Chem. Phys. {\bf 105}, 9299 (1996).
\bibitem{Dellago1998a} C. Dellago, P. G. Bolhuis, F. S. Csajka, and D. Chandler, J. Chem. Phys. {\bf 108}, 1964 (1998).
\bibitem{Passerone2001} D. Passerone and M. Parrinello, Phys. Rev. Lett. {\bf 87}, 108302 (2001).
\bibitem{Fujisaki2010} H. Fujisaki, M. Shiga, and A. Kidera, J. Chem. Phys. {\bf 132}, 134101 (2010).
\bibitem{Lee2017} J. Lee, I.-H. Lee, I. Joung, J. Lee, and B. R. Brooks, Nat. Commun. {\bf 8}, 15443 (2017).
\bibitem{VanErp2003} T. S. van Erp, D. Moroni, and P. G. Bolhuis, J. Chem. Phys. {\bf 118}, 7762 (2003).
\bibitem{Moroni2004} D. Moroni, P. G. Bolhuis, and T. S. van Erp, J. Chem. Phys. {\bf 120}, 4055 (2004).
\bibitem{Vlugt2001} T. J. H. Vlugt and B. Smit, Phys. Chem. Comm. {\bf 4}, 11(2001).
\bibitem{Borrero2016} E. E. Borrero and C. Dellago, Eur. Phys. J-Spec. Top. {\bf 225}, 1609 (2016)
\bibitem{Bolhuis2018a} P. G. Bolhuis and G. Cs{\'{a}}nyi, Phys. Rev. Lett. {\bf 120}, 250601 (2018).
\bibitem{Onsager1953} L. Onsager and S. Machlup, Phys. Rev. {\bf 91}, 1505 (1953).
\bibitem{Parrinello1984} M. Parrinello and A. Rahman, J. Chem. Phys. {\bf 80}, 860 (1984).
\bibitem{SM} See Supplemental Material for more details about the simulation setup, additional results and discussions. Includes Refs.~\cite{jorgensen1983,Chandler1987,Hockney,Martina1992,Putrino2000,Tribello2014,Kapil2016}.
\bibitem{Calhoun1996} A. Calhoun, M. Pavese, and G. A. Voth, Chem. Phys. Lett. {\bf 262}, 415 (1996).
\bibitem{Plimpton1995} S. Plimpton, J. Comput. Phys. {\bf 117}, 1 (1995).
\bibitem{Bonomi2009} M. Bonomi, and A. Barducci, and M. Parrinello, J. Comput. Chem. {\bf 30}, 1615 (2009).
\bibitem{Branduardi2012} D. Branduardi, G. Bussi, and M. Parrinello, J. Chem. Th. Comput. {\bf 8}, 2247 (2012).
\bibitem{tiwary2015} P. Tiwary and M. Parrinello, J. Chem. Phys. B {\bf 119}, 736 (2015).
\bibitem{Mones2016} L. Mones, N. Bernstein, and G. Cs\'ani, J. Chem. Th. Comput. {\bf 12}, 5100 (2016).
\bibitem{Marinova2019} V. Marinova and M. Salvalaglio, J. Chem. Phys. {\bf 151}, 164115 (2019).
\bibitem{invernizzi2020} M. Invernizzi and M. Parrinello, J. Phys. Chem. Lett. {\bf 11}, 2731 (2020).
\bibitem{giberti2020} F. Giberti, B. Cheng, G. A. Tribello, and M. Ceriotti, J. Chem. Th. Comput. {\bf 16}, 100 (2020).
\bibitem{Miller1974} W. H. Miller, J. Chem. Phys. {\bf 61}, 1823 (1974).
\bibitem{Miller1983} W. H. Miller and S. D. Schwartz and J. W. Tromp, J. Chem. Phys. {\bf 79}, 4889 (1983).
\bibitem{Barducci2008} A. Barducci, G. Bussi, and M. Parrinello, Phys. Rev. Lett. {\bf 100}, 020603 (2008).
\bibitem{Dama2014} J. F. Dama, M. Parrinello, and G. A. Voth, Phys. Rev. Lett. {\bf 112}, 240602 (2014).
\bibitem{Weismiller2010} M. R. Weismiller, A. C. T. van Duin, J. Lee, and R. A. Yetter, J. Phys. Chem. A {\bf 114}, 5485 (2010).
\bibitem{Valsson2016} O. Valsson, P. Tiwary, and M. Parrinello, Annu. Rev. Phys. Chem. {\bf 67}, 159 (2016).
\bibitem{Rosa-Raices2019} J. L. Rosa-Raı́ces, B. Zhang, and T. F. Miller, J. Chem. Phys. {\bf 151}, 164120 (2019).
\bibitem{Car1985} R. Car and M. Parrinello, Phys. Rev. Lett. {\bf 55}, 2471 (1985).
\bibitem{jorgensen1983} W. L. Jorgensen, J. Chandrasekar and J. D. Madura, J. Chem. Phys. {\bf 79}, 926 (1983).
\bibitem{Chandler1987} D. Chandler, {\it Introduction to Modern Statistical Mechanics} (New York, 1987).
\bibitem{Hockney} R. W. Hockney and J. W. Eastwood, {\it Computer Simulation Using Particles } (New York, 1989).
\bibitem{Martina1992} G. J. Martyna, M. L. Klein, and M. Tuckerman, J. Chem. Phys. {\bf 97}, 2635 (1992).
\bibitem{Putrino2000} A. Putrino, D. Sebastiani, and M. Parrinello, J. Chem. Phys. {\bf 113}, 7102 (2000).
\bibitem{Tribello2014} G. A. Tribello, M. Bonomi, D. Branduardi, C. Camilloni, and G. Bussi, Comput. Phys. Commun. {\bf 185}, 604 (2014).
\bibitem{Kapil2016} V. Kapil, J. Behler, and M. Ceriotti, J. Chem. Phys. {\bf 145}, 234103 (2016).
\end{thebibliography}
\end{document}